\documentclass[letterpaper,dvipsnames,11pt]{article}
\pdfoutput=1
\usepackage{jheppub}
\usepackage{amsmath}
\usepackage{graphicx}

\usepackage{soul}
\usepackage{array}
\usepackage{xcolor}

\usepackage{slashed}
\usepackage{afterpage}
\usepackage{appendix}
\usepackage{multirow}
\usepackage{float}
\usepackage{mathtools}
\usepackage{tensor}
\mathtoolsset{centercolon}

\usepackage{multirow}

\newcommand{\MPl}{M_{\rm Pl}}
\newcommand{\pd}{\partial}

\newcommand{\QBH} {Q_{\rm BH}}

\definecolor{darkgreen}{RGB}{50,150,0}

\preprint{CERN-TH-2025-256}

\title{Charged black holes in the 1/N expansion}

\author[a,b,c]{Georges Obied} 
\author[a,d]{and Mario Reig}

\emailAdd{gobied@uchicago.edu}
\emailAdd{mario.reig.lopez@cern.ch}

\affiliation[a]{Rudolf Peierls Centre for Theoretical Physics, 
University of Oxford, Parks Road, Oxford OX1 3PU, United Kingdom}
\affiliation[b]{Leinweber Institute for Theoretical Physics \& Enrico Fermi Institute \& Kadanoff Center for Theoretical Physics, University of Chicago, Chicago, IL 60637, USA}
\affiliation[c]{Kavli Institute for Cosmological Physics, University of Chicago, Chicago, IL 60637, USA}
\affiliation[d]{Theoretical Physics Department, CERN, 1211 Geneva 23, Switzerland}

\abstract{We study some implications of $SU(N)\times U(1)$ theories coupled to gravity in the large-$N$ limit.  We find that in theories  with quarks transforming as $q\sim (\mathbf{N},1)$, black holes (BH) with charge larger than a critical value approach extremality $(M_{\rm BH}=\sqrt{2}\MPl Q_{\rm BH})$ as they evaporate. This occurs because BHs in this theory can only lose charge by emitting baryons, a process suppressed by a combinatoric factor $e^{-N\log N}$. Once extremality is reached the BH evolution halts for exponentially long times, $\tau_{\rm BH}\gtrsim e^{N\log N}$, without being protected by any symmetry. At times $t\gg\tau_{\rm BH}$ the BH forms a cloud of deconfined quarks around it, a \textit{quark corona}. The quarks are connected to the horizon by strings extending throughout distances that can be much larger than the inverse of the confinement scale, $\Lambda_N^{-1}$. Strikingly, due to energy conservation, these long strings cannot break even for light quarks. These effects can be further enhanced in clockwork-like theories based on Yang-Mills sectors of the type, $SU(N_1)\times SU(N_2)\times ...\times U(1)$. As an application to phenomenology, we give examples where BHs as light as one gram survive until today without evaporating, opening up regions of parameter space for primordial BH dark matter previously excluded by Hawking evaporation. This result relies only on large-$N$ combinatorics and is independent of the radiation mechanism from the BH, particle masses, and confinement scale. The theories we discuss pass non-trivial tests imposed by Swampland conjectures, including completeness of the spectrum as well as the weak gravity conjecture.
}

\begin{document}

\maketitle

\section{Introduction}
Black holes (BHs) are among the most fascinating systems in physics. In recent decades BHs have been studied in a variety of contexts, ranging from holography~\cite{Aharony:1999ti} and string theory~\cite{Strominger:1996sh} to astrophysics~\cite{haardt2016astrophysical} and as dark matter candidates~\cite{Carr:2020xqk}. 
Hawking discovered that BHs emit thermal radiation with temperature $T_H=\frac{\kappa}{2\pi}$, where $\kappa$ is the surface gravity~\cite{Hawking:1975vcx}. Subsequently, Gibbons~\cite{Gibbons:1975kk} showed that for large charged BHs, Hawking emission of charged particles can be thought of as Schwinger emission from the BH electric field. Taking this viewpoint, Hiscock and Weems \cite{Hiscock:1990ex} studied quantitatively how Schwinger emission of charged particles determine the evolution of electrically charged (Reissner-Nordstr\"om) BHs. More recently, the full evolution of charged BHs including quantum effects has been presented in~\cite{Brown:2024ajk}, indicating that neutral Hawking radiation is dramatically slowed down as one approaches extremality. It is also possible that BHs have magnetic charges~\cite{Maldacena:2020skw}, leading to very rich phenomenology~\cite{Bai:2020spd,Araya:2020tds,Diamond:2021scl}. In the context of the Standard Model (SM), this may lead to the formation of a \textit{hairy} corona where the EW symmetry is restored. This corona may extend throughout distances much beyond the EW scale. 

If there exist hidden sectors that do not communicate efficiently with the SM (e.g. only via gravitational interactions), BHs can be colder than naively expected. A BH that is extremal with respect to a dark $U(1)$ gauge symmetry has zero temperature. For a visible sector observer this fact may be rather confusing because according to its mass the BH should be much hotter than it actually is or should even have evaporated by today. Similar ideas have been previously considered in the context of dark QED in~\cite{Bai:2019zcd}, where it is shown that having sufficiently heavy fermions  (close to the Planck scale) with dark $U(1)$ charge can lead to stable, light primordial BH dark matter. It is worth mentioning that the existence of these non-supersymmetric exactly extremal solutions with zero temperature has been challenged recently based on theoretical grounds (see~\cite{Turiaci:2023wrh} and references therein), but for the purpose of this work we will adopt them as valid systems. This is because, for the  phenomenological applications that will be discussed, it is sufficient that the BH only be close to extremality rather than exactly extremal.

Confining $SU(N)$ theories in the limit of large number of colours~\cite{tHooft:1973alw,Witten:1979kh} present very interesting features. Mesons are weakly coupled point particles, becoming non-interacting and stable in the limit $N\rightarrow \infty$, while baryons behave as solitonic objects analogous to magnetic monopoles~\cite{Witten:1979kh}. Indeed, baryons have stronger couplings and are objects with finite size, set by the $SU(N)$ confinement scale $\Lambda_N$. This means that the product of the baryon mass and size grows with the number of colours, $M_{\rm B}R_{\rm B}\sim N$. 

In this article we consider a modification of a minimal (non-supersymmetric) large-$N$ theory by adding a QED-like interaction, $SU(N)\times U(1)$, (which throughout the article will be called EM, although we assume it is a hidden sector) and couple it to gravity. We will assume that the spectrum contains a single Dirac fermion transforming as $q\sim (\mathbf{N},1)$. (We will refer to this particle as a \textit{quark}, although we note that it is not related to one of the SM fermions with $SU(3)_c$ charge.) In the limit of a large number of colours we will see that this class of theories have very interesting implications for BH physics. BH solutions with colour charge have been studied in the past~\cite{PhysRevLett.64.2844,Volkov:1998cc} as well as more recently in the context of primordial BHs with SM QCD colour charge~\cite{Alonso-Monsalve:2023brx}. In our case, we will instead be interested in Reissner-Nordstr\"om(-like) solutions where, asymptotically, the BH has a dark $U(1)$ charge.

The Yang-Mills (YM) $SU(N)$ theory confines at an energy scale $\Lambda_N$ below which coloured particles are bound in hadronic colour singlets. We assume a picture of confinement where the confining string induces a linear potential between a quark-antiquark pair, $V(r)\sim \Lambda_N^2 r$, with $\Lambda_N^2$ being the string tension. In the large-$N$ limit, the confining strings are non-interacting \cite{Athenodorou:2021qvs} and one can understand them as being approximately equivalent to $U(1)$ flux tubes that confine magnetic monopoles~\cite{Nambu:1974zg}. Indeed, some of our findings can also be mimicked by a theory based on spontaneously broken $SU(2)\rightarrow U(1)$ by a Higgs triplet, and $U(1)\rightarrow 0$ by a different Higgs triplet. In this case one has (solitonic) monopoles confined by strings, and BHs with magnetic charge will have similar behaviour to the ones we describe here. Indeed, this latter setup is similar in spirit to the one studied in~\cite{Maldacena:2020skw} in the context of the SM.

Some of the findings in this article are summarised by the following gedanken experiment. Let us imagine that a large-$N$ baryon -- that is, a state composed of $N$ quarks with totally antisymmetric contraction of colour indices, $B=\epsilon^{i_1...i_N} q^{i_1}...q^{i_N}$ -- falls into an uncharged BH. In this process, the BH gains an electric charge equal to the baryon charge, given by $Q_{\rm BH}=eN$, where $e$ is the $U(1)$ gauge coupling. The subsequent evolution of the BH occurs as follows. While neutral Hawking radiation, which includes photons, gravitons and possibly mesons follows the standard emission rate for a given BH temperature~\cite{Hawking:1975vcx}, Hawking and Schwinger emission of charged states (in this case baryons) receive an additional exponential suppression. This is reminiscent of the interesting observation by Witten~\cite{Witten:1979kh} that the scattering rate of the process $e^+e^-\rightarrow B \bar{B}$ is suppressed in the large-$N$ limit\footnote{Note that in our case there are no leptons, but processes like $\pi \pi \rightarrow B\bar{B}$ are similarly suppressed.} as
\begin{equation}\label{eq:baryon_rate_suppressed}
    \Gamma\propto e^{-N\log N}\,.
\end{equation}
We note that this exponential suppression is independent of the quark masses and the confinement scale, and operates even if the temperature of the BH is comparable to (or above) the baryon mass. Intuitively, this suppression occurs because producing baryons on-shell requires $N$ quarks with the correct combination of colour charges, so that their totally antisymmetric contraction gives a colour singlet. The probability that this occurs is $p\sim \left ( \frac{1}{N} \right )^N$. Hence, Hawking or Schwinger emission of baryons is suppressed by a factor comparable to that in Eq.~\eqref{eq:baryon_rate_suppressed}.  Quite surprisingly, BHs in this kind of theory can lose mass efficiently -- via standard Hawking evaporation -- but not electric charge. 

Together with confinement, which precludes the possibility of emitting individual free quarks, this simple observation implies that in $SU(N)\times U(1)$ theories charged BH have very interesting and novel properties. We find that above a certain critical charge that will be determined later, a BH with arbitrary charge and mass approaches extremality, $M_{\rm BH}=\sqrt{2}\MPl Q_{\rm BH}$, without losing charge. After reaching extremality, the BH evaporation halts because the temperature vanishes, $T_H=0$, and the BH can only emit charge via the Schwinger production of baryons. As argued above, however, this process is exponentially slow. The characteristic time for the BH to emit a sizeable fraction of the $U(1)$ charge it has is $\tau_{\rm BH} \gg e^{N\log N}$ (normalised to the characteristic timescale, nearly given by the electric field as $\sim E^{-1/2}$). The subsequent evolution tracks the extremality condition -- the emission of charge via baryon production is necessarily followed by more efficient neutral Hawking radiation that brings the BH close to extremality. These are examples of extremal black holes that can persist in our universe, providing a counter-example to the bounds of \cite{Coviello:2025slv} obtained in the context of Maxwell-like theories with point-like charged particles.

Eventually, the BH reaches a critical mass $M_{\rm crit}$ and charge $Q_{\rm crit}$ for which the electric field is comparable to the string tension, $eE \sim \Lambda_N^2$. After this, the BH enters a qualitatively new phase in which quark--antiquark pairs attached by strings are nucleated via Schwinger pair production. While the antiquark falls into the BH (effectively lowering its charge), the quark is suspended outside the horizon due to the balance between attractive forces, due to gravity and the string tension, and the repulsive electromagnetic force. We will show that this leads to a cloud of deconfined quarks that can extend throughout distances $l$ much larger than the confinement scale, $l\gg \Lambda_N^{-1}$. We study the properties of this corona in the static limit in which quark production is slow. The qualitative evolution of this new kind of BH is summarised in Fig.~\ref{fig:BH_corona-parameter_space}.

The BH solutions that we find may have interesting implications for phenomenology. Due to their exponentially long life-time, these charged BH could compose the observed abundance of DM even if they are much lighter than the standard evaporation bound~\cite{Carr:2020xqk}. As an example, we will see that the for $N\sim 30$ a BH can be cosmologically stable even if it is as light as a $M_{\rm BH}\sim 1$ gram. We note that this is due to the fact that the BH survives as an extremal BH for very long times $\tau \sim G^2M_{\rm BH}^3e^{N\log N}$. We also note that this stability does not come from symmetries but, instead, comes from combinatoric suppression in the limit of a large number of colours. Thinking about the baryons as solitonic objects of $SU(N)$, one can argue that the suppression comes from the fact that the emission of such solitons is exponentially suppressed, $e^{-S}$, with $S$ being a large action (see \cite{Johnson:2018gjr} for a related discussion)\footnote{This is why solitonic monopoles confined by magnetic flux also would give rise to similar situation in the case of theories with spontaneous breaking $SU(2)\rightarrow U(1)\rightarrow 0.$}. Despite this exponentially long life-time, we will show that the large-$N$ theories we consider are compatible with the theoretical restrictions imposed by different swampland conjectures~\cite{Vafa:2005ui,Brennan:2017rbf,Palti:2019pca,vanBeest:2021lhn}, including completeness of the spectrum~\cite{Heidenreich:2021xpr}, the species bound~\cite{Dvali:2007hz,Dvali:2007wp} as well as the weak gravity conjecture~\cite{Arkani-Hamed:2006emk}. 

Recently, it has been also argued that the Hawking evaporation process could be slowed for neutral BHs due to the so-called memory burden effect~\cite{Dvali:2018xpy,Dvali:2024hsb,Alexandre:2024nuo}. The key aspect of this effect is that BHs store a large amount of information which slows, or even prevents, their decay (although that view has been recently challenged in~\cite{Montefalcone:2025akm}, where the authors argue that the memory burden protection might only hold in some cases). While the mechanisms that we study here have similar consequences -- that is, a confining large-$N$ theory may be used to stabilise light PBHs -- we note that the origin of the BH stability is different from the memory burden. In the cases we study, as we will see, the stability will ultimately come from the difficulty in having baryon emission from a charged BH. It would be nice to find more explicit connections between both mechanisms. 

The article is organised as follows. To gain some intuition, in section \ref{sec:flat_space}, we describe the behaviour of $SU(N)\times U(1)$ theories using a capacitor in flat space as an example. This allows us to illustrate how in this class of theories, large electric fields ($eE\gtrsim \Lambda_N^2$) can deconfine quark-antiquark pairs attached by a string. After that, in sections \ref{sec:including_gravity} and \ref{sec:dynamics}, we describe the theory including gravity. First, we show how BHs with mass and charge above the critical values are exponentially long-lived. We also show that there exist solutions where quarks are suspended outside the horizon and solve the Einstein equations to find the metric in the limit of slow quark-antiquark production. After that, we describe the evolution of the system.
In section \ref{sec:swamp} we discuss the compatibility of long-lived extremal BHs with the Swampland conjectures.  In section~\ref{sec:pheno} we discuss some of the phenomenological implications, including how a large-$N$ theory can stabilise \textit{ultralight} PBHs. Finally, we conclude in section~\ref{sec:conclusion}.

\section{Quark deconfinement in an electric field: the capacitor case}
\label{sec:flat_space}
In preparation for the following discussion on the case of a charged BH, it is instructive to start with a more familiar example: a capacitor in flat space. Let us consider a theory based on $SU(N)\times U(1)$ with quarks with mass $m_q$ transforming as
\begin{equation}
    q\sim (\mathbf{N},1)\,.
\end{equation}
Let us assume that inside the capacitor there is a constant electric field, $E$, and study quark-antiquark pair creation through the Schwinger effect. Unlike standard pair production of $e^+e^-$ in EM, in a theory where coloured particles (quarks) are confined by a string there is a minimum electric field for this process to occur. Pair production of quark-antiquarks pairs only takes place for large enough electric fields,
\begin{equation}
    eE>\Lambda_N^2\,.
\end{equation}
This is due to the string tension between confined particles and is analogous to the magnetic monopole pair production considered in \cite{Hook:2017vyc}. In this section we show that, for $eE>\Lambda_N^2$ (where $e$ is the $U(1)$ gauge coupling), the confining strings do not break and instead we have far separated quark-antiquarks independently of the quark mass, $m_q$. In other words, once Schwinger pair production is allowed, it is energetically preferred to have long strings of length $l\gg \Lambda^{-1}_N$ extended throughout the capacitor instead of shorter strings, $l\lesssim \Lambda^{-1}_N$. Therefore, we find that electric fields `deconfine' quarks in a way similar in spirit to magnetic fields restoring the EW symmetry \cite{Ambjorn:1989bd,Ambjorn:1989sz,Ambjorn:1992ca,Tornkvist:1992kh,Chernodub:2012fi}. We note, however, that this analogy is only accurate in the large-$N$ limit where the strings are non-interacting \cite{Athenodorou:2021qvs}.

Schwinger pair-production is equivalent to tunneling through a barrier. In the case of interest, the instanton (critical bounce) action can be estimated to be \cite{Hook:2017vyc}\footnote{In order to trust this estimates, we assume to be in the regime $E<E_c\lesssim \frac{m_q^2}{eq_e}$, with $e\sim 1/\sqrt{N}$ the gauge coupling and $q_q=1$ the quark electric charge. We therefore assume that the quarks have a mass $m_q^2>(eE-\Lambda_N^2)$ so that we keep the Schwinger production under control. One might be interested in dropping this assumption and study Schwinger in the presence of massless fermions.}
\begin{equation}\label{eq:Schwinger_rate_quarks}
    S=\frac{m_q^2}{eE-\Lambda_N^2}\,.
\end{equation}
The distance between the quark and antiquark at the time of nucleation can be easily estimated assuming conservation of energy in the tunneling process,
\begin{equation}
    \Delta x=\frac{2m_q}{eE-\Lambda_N^2}\,.
\end{equation}
Assuming confinement by a linear potential, $V(r)=\sigma_Nr\sim \Lambda_N^2 r$, the potential energy for a quark-antiquark pair separated by a distance $\Delta x$ inside an electric field $E$ is
\begin{equation}
    V(\Delta x)= 2m_q-\Delta x(eE-\Lambda_N^2)\,.
\end{equation}
The second term tells us that moving the quark--antiquark farther apart lowers the potential energy of system (i.e. there is a force driving the system towards a well-separated $q \bar{q}$ pair).

Consider now a capacitor of size $d\gg \Lambda_N^{-1}$ and constant electric field 
\begin{equation}
    m_q^2>eE-\Lambda_N^2> 0\,,
\end{equation}
so that Schwinger production is slow enough to be under control. After nucleation,  the quark and antiquark are accelerated towards  opposite walls of the capacitor leaving a \textit{gas} of long strings inside the capacitor. Strikingly, as long as $eE>\Lambda_N^2$, the string will not break independently of the quark mass. 

\begin{figure*}[t]
	\centering
	\includegraphics[width=0.6\textwidth]{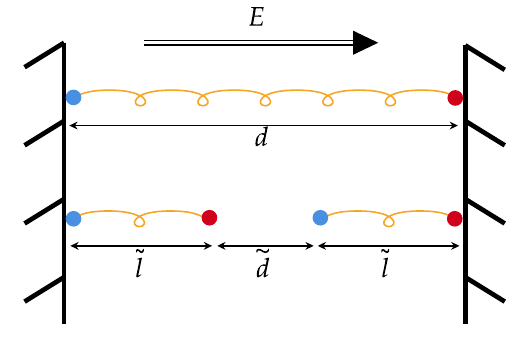}
		\caption{Two different configurations with strings of different lengths used to compare their (free) energy. The top one corresponds to unbroken long strings with energy $\mathcal{E}_1$. The bottom is a configuration after nucleating a $q\bar{q}$ pair on the flux. This configuration has energy $\mathcal{E}_2$. As discussed in the main text, we always have $\mathcal{E}_2 > \mathcal{E}_1$ and the breaking of flux tubes by quark nucleation is energetically forbidden.}
	\label{fig:capacitor}
\end{figure*}

The configuration where the size of the string is $l\gg \Lambda_N^{-1}$ can be shown to be the minimum (free) energy state, indicating that this is the equilibrium configuration for $eE>\Lambda_N^2$.
To see this, let us take a string of length $l\sim d$, whose endpoints are stuck at the oppositely charged walls of the capacitor. The total energy of the system with a constant electric field $E$ is
\begin{equation}
\mathcal{E}_1=d\Lambda_N^2 + 2m_q+\mathcal{E}_0\,,
\end{equation}
where $\mathcal{E}_0$ is the energy stored in the constant electric field. If the string breaks (around the mid part, for simplicity) into two smaller strings of size $\tilde l$, two endpoints are still stuck at the walls of the capacitor while the other two (quarks and antiquarks) are under the influence of the electric field. (See Fig.~\eqref{fig:capacitor}.) Let us assume that the distance between the free endpoints is given by $\tilde{d}$. The total energy of this second configuration is:
\begin{equation}
 \mathcal{E}_2=\tilde{d}eE+2\tilde{l}\Lambda_N^2+4m_q+\tilde{\mathcal{E}}_0\,,
\end{equation}
where $\tilde{d}eE$ is the energy of the \textit{dipole} due to the \textit{free} endpoints in an electric field $E$. Since $d=2\tilde{l}+\tilde{d}$ and $eE>\Lambda_N^2$, the energy of the system with broken, shorter strings and free endpoints is larger than the energy of the long string of length $d$:
\begin{equation}
    \Delta \mathcal{E}=\mathcal{E}_2-\mathcal{E}_1=2m_q+\tilde{d}(eE-\Lambda_N^2)>0\,.
\end{equation}
Note that this does not depend on the size of the capacitor $d$, so it holds even if $d\Lambda_N^2\gg m_q$. This demonstrates that, in a large-$N$ theory, a constant electric field $eE>\Lambda_N^2$ is able to deconfine quarks and confirms that with long strings are energetically preferred\footnote{At zero temperature the Helmholtz free energy coincides with the internal energy. The  state minimizing the energy $\mathcal{E}$ -- that is, a state with long strings stretched between the walls of the capacitor -- is the equilibrium state. Moreover, in the particular case $m_q^2>eE\sim \Lambda_N^2$, quantum fluctuations are very suppressed and we can understand the system using classical physics, although we do not require that case in general.}. Note that we have neglected backreaction into the electric field of the capacitor, $\mathcal{E}_0=\tilde{\mathcal{E}}_0$. Including backreaction effects will only have positive contributions to $\Delta \mathcal{E}$ resulting in the state with long strings more energetically preferred.

After nucleating many pairs connected by long strings, the system will eventually reach equilibrium. This is achieved as the electric field is lowered and approaches
\begin{equation}
    eE\sim \Lambda_N^2\,,
\end{equation}
that is, when the instanton action diverges and Schwinger production of $q\bar{q}$ pairs stops. 

\subsection{Pair production of baryons and discharge of the electric field}
A capacitor full of long strings and electric field $eE\sim \Lambda_N^2$ will survive for exponentially long times. Indeed, we have seen that $q\bar{q}$ production stops and the long strings cannot break. The electric field, however, eventually discharges due to Schwinger pair production of colour singlet baryon antibaryon pairs, $B\bar{B}$. In large-$N$ theories the baryon state, $B$, is composed of $N$ quarks with totally antisymmetric contraction of their colour index. Their quantised charge is  $q_{\rm B}=N$, and their mass is roughly given by
\begin{equation}
    M_{\rm B}\sim N(m_q+\Lambda_N)\,.
\end{equation}

We now consider pair production of baryons $B\bar{B}$ in the capacitor. In addition to the usual Schwinger (instanton-like) suppression, in the large-$N$ limit the production of baryons has an additional exponential suppression factor. This can be intuitively understood with simple combinatorics arguments. In order to form a baryon, one requires to have $N$ quarks with the appropriate colour charges. The probability for this to happen in a single fluctuation is roughly
\begin{equation}
    p\sim \left (\frac{1}{N}\right )^N = e^{-cN}\,,\,\text{ with }c=\log N\,.
\end{equation}
As mentioned in the introduction, this exponential suppression for the Schwinger production rate of baryons is analogous to that in other processes, such as the $e^+e^-\rightarrow B\bar{B}$ considered by Witten \cite{Witten:1979kh}. 

The baryon-anti baryon production rate is  suppressed as:
\begin{equation}\label{eq:baryon_prod_rate_capacitor}
\Gamma_{B\bar{B}}\propto e^{-S_{\rm Sch.}}e^{- N\log N}\,\,,\,\,\text{with:}\,\, S_{\rm Sch.}\sim \frac{M_{\rm B}^2}{q_{\rm B}eE}\sim \frac{(m_q+\Lambda_N)^2}{\Lambda_N^2}N\,.
\end{equation}
Therefore, the lifetime for the equilibrium state of a capacitor with long strings is expected to be at least of order
\begin{equation}
    \tau \sim \left ( eE\right )^{-1/2} e^{\frac{N(m_q+\Lambda)^2}{eE}+N\log N}\,.
\end{equation}
In the following subsection~\ref{sec:N_individual_quarks}, we explain why this conclusion is still valid even if one considers assembling the baryon by independently producing $N$ individual quark-antiquark pairs.

After an exponentially long time, multiple $B\bar B$ production results into electric field discharge, and we eventually reach a phase with
\begin{equation}
    eE\lesssim \Lambda_N^2\,.
\end{equation}
The system is now classically unstable, since the neat force between $q\bar q$, is now attractive -- the string tension overcomes the force exerted by the external, constant electric field. In a short period of time, $t\lesssim d$, the $q\bar q$ pairs connected by long strings will shrink to sizes $l\sim \Lambda^{-1}_N$, that is mesons. This means that the system has decayed into a capacitor with $eE\lesssim \Lambda_N^2$ and a \textit{gas} of mesons. The mesons -- which in the case of a single flavour is an $\eta^\prime$-like meson -- may decay into photons via the anomaly. 

Finally, a relevant question is whether long strings can break through quark-antiquark production due to the chromo-electric field along the string. First, we note that as shown above string breaking is not energetically allowed for $eE>\Lambda_N^2$, irrespective of the quark mass. On the other hand, once $eE\lesssim \Lambda_N^2$, the dynamics of string breaking depends on $m_q$. For heavy quarks $m_q>\Lambda_N$, in analogy to the \textit{quirk} scenario \cite{Kang:2008ea}, the timescale associated to the breaking reads
\begin{equation}
    \tau_{q\bar{q}} = \Gamma^{-1}\sim \Lambda_N^{-1}e^{m_q^2/\Lambda_N^2}\,.
\end{equation}
This timescale can be extremely long  even for mild hierarchies between the string tension and $m_q^2$.  On the other hand, the timescale to reach the equilibrium configuration where the string connecting both $q$ and $\bar{q}$ shrinks to $l\sim \Lambda_N^{-1}$ is roughly $t\lesssim d$, that is the size of the capacitor. For light quarks $m_q\ll \Lambda_N$, the string breaking may be fast enough to compete with the classical contraction of the $q\bar{q}$ pair.

\subsubsection{Production of $N$ individual quark pairs}\label{sec:N_individual_quarks}

Before closing this section, we discuss the production rate of $N$ individual quark-antiquark pairs and compare it to producing a baryon-anti baryon pair in a single, large quantum fluctuation. While the rate of producing $N$ individual pairs scales as $\Gamma \propto e^{-NS_{q\bar q}}$, which seems naively to be less suppressed than the production of a $B\bar B$ pair (see Eq.~\eqref{eq:baryon_prod_rate_capacitor}), we note that for this process to lead to a production of baryons, the $N$ pairs should be nucleated in the same region of the capacitor. The probability that $N$ pairs are nucleated in the same region of volume $\Lambda_N^{-3}$ scales as $p\sim (V_{\rm tot}\Lambda_N^3)^{-N}$. This leads to an additional suppression factor for the rate of interest that scales as
\begin{equation}
    e^{-N\log (V_{\rm tot}\Lambda_N^3)}\,,
\end{equation}
with $V_{\rm tot}$ being the volume of the capacitor. When the total volume of the capacitor is $V_{\rm tot}\gg \Lambda_N^{-3}$, this factor makes the production of $N$ quark pairs more suppressed than the baryon-antibaryon process.

\section{Large-$N$ black holes}
\label{sec:including_gravity}
In the previous section we discussed the behaviour of a large-$N$ theory in the presence of a background electric field in flat space (i.e. a capacitor). We found that, for a particular range of parameters, we produce quark--antiquark pairs attached by confining strings of length $d \gg \Lambda_N^{-1}$. In this section, we will extend our discussion to include gravity. We will couple the large-$N$ $SU(N)\times U(1)$ theory to gravity and study Reissner-Nordstrom (RN) black holes charged under $U(1)$. Similar to the capacitor, a large electric field near the black hole horizon can lead to quark--antiquark pair production via the Schwinger effect. The electric field stretches the confining flux tubes so that the antiquark falls into the black hole while the quark is held at a macroscopic distance~$\gg \Lambda_N^{-1}$ from the BH horizon. We show that this process leads to the formation of a cloud or corona of deconfined quarks around the BH. Roughly speaking, this BH provides an electric counterpart to the magnetically charged black holes with an electroweak corona discussed in \cite{Maldacena:2020skw}. These black holes present new features that we will discuss below.
We will also show that thanks to the spherical symmetry of these BH one can solve the Einstein equations and obtain a metric for this spacetime.

\subsection{Review of charged BH}

Let us start with a brief review of the RN solution to set notation. We will be interested  in a theory containing gravity and electromagnetism (more rigurously, an unbroken $U(1)$ gauge theory) with an action
\begin{align}
S = \int d^4x \sqrt{-g} \left(\frac{\MPl^2}{2}\mathcal{R} - \frac14 F^2 + \ldots \right)\,,
\end{align}
where we are showing only the Einstein-Hilbert and Maxwell terms that are important for the RN solution that we will consider. The metric of these black holes can be found in any standard textbook on general relativity (such as \cite{Misner:1973prb}) and takes the form
\begin{align}
\label{eq:RNmetric}
ds^2 = -U(r) dt^2 + U(r)^{-1} dr^2 + r^2 d\Omega^2\,,\\
U(r) = 1 - \frac{2GM_{\rm BH}}{r} +  \frac{G Q_{\rm BH}^2}{4\pi r^2}\,,
\end{align}
where $8\pi G  = \MPl^{-2}$, $M_{\rm BH}$ is the black hole mass and $Q_{\rm BH}=eQ_{\rm quant}$ is the black hole charge (which includes $U(1)$ the gauge coupling and the quantized part). Just like their uncharged counterparts, these black holes have a singularity at $r=0$ which is cloaked by a horizon. In fact, these black holes have two horizons at the coordinate radii:
\begin{align}
\label{eq:RNhorizons}
r_\pm = GM_{\rm BH} \pm \sqrt{(GM_{\rm BH})^2 - \frac{G Q_{\rm BH}^2}{4\pi}}.
\end{align}
Physical solutions are sub-extremal black holes for which the horizons exist and the singularity is hidden from far away observers. (The existence of non-supersymmetric extremal BH solutions with zero temperature has been recently challenged in \cite{Turiaci:2023wrh}.). These have $M_{\rm BH}^2 \geq 8 \MPl^2 Q_{\rm BH}^2$. The above RN solution is supported by electromagnetic fields with gauge potential
\begin{align}
\label{eq:EMfields}
A = -\frac{Q_{\rm BH}}{4\pi r} dt\,.
\end{align}
Finally, these black holes radiate with a Hawking temperature given by
\begin{align}
\label{eq:RNtemperature}
T_H = \frac{r_+ - r_-}{4\pi r_+^2}\,.
\end{align}
For observers at a fixed coordinate radius outside the horizon, the temperature is increased by a factor of $U(r)^{-1/2}$. Freely-falling observers, however, experience a temperature similar in magnitude to $T_H$ \cite{Brynjolfsson:2008uc}. Finally, and without loss of generality, we will assume that the black hole has positive charge (i.e. $Q_{\rm BH}>0$) in the discussion below. 

\subsection{New BH solutions}
As we saw in the previous section, quark-antiquark production can happen in the presence of strong electric fields even when the non-Abelian theory is confined at long distances. Similar to the capacitor, the electric field of a charged black hole spontaneously produces quark--antiquark pairs   attached by a confining string. The electric field of the black hole stretches this string causing the antiquark (with electric charge opposite that of the BH) to fall behind the horizon while the quark remains suspended outside the BH. Heuristically, the quark is held in place by a balance of the attractive gravitational and tension forces against electromagnetic repulsion. This process also gives the black hole an anticolour charge (i.e. the charge of the antiquark). Altogether, the production of quark-antiquark pairs leads to the population of a quark corona around the black hole. 

In this section we focus on static, equilibrium solutions for this classical string suspended outside a charged black hole and postpone the description of the evolution and dynamical production of quark-antiquark pairs for later sections. We note that production of long confining strings attached to a quark can happen for any charged black hole, as long as there is a region with electric field satisfying $eE > \Lambda_N^2$, cf. the discussion in the previous section. That said, for simplicity, we will specialise our discussion below to the case of extremal black holes  despite the fact that this process is more general, i.e. happens for sub-extremal black holes, too. Of course, for a given mass $M_{\rm BH}$, extremal black holes have the largest electric fields near their horizon and satisfy the condition $eE > \Lambda_N^2$ more readily. One may wonder about the possibility of breaking the confining string due to the nucleation of a quark--antiquark pair. However, as argued above (see sec.~\ref{sec:flat_space}), in regions where the electric field satisfies $eE > \Lambda_N^2$, the process of string breaking is forbidden by energetics. In the case of BHs, in addition, the process of string breaking would lead to the emission of neutral mesons which, for the case of extremal BHs, would lead to super-extremal BHs. We will show below that the string, and therefore the quark corona, only exists in regions where $eE > \Lambda_N^2$, so its breaking is not relevant for our discussion.

\subsubsection{Quarks suspended outside black holes}
Let us now describe the classical solution corresponding to a single flux tube emanating from a black hole with a charged quark at its end. We will treat the flux tube and quark in the probe approximation which is to say that we neglect their back-reaction on the black hole geometry. As such, we will assume the spacetime geometry is given by the metric in~\eqref{eq:RNmetric}. The system we are considering is then described by the following degrees of freedom. First there is a quark worldline 
\begin{align}
    q^\mu(\lambda) = \left(t(\lambda), r(\lambda), \frac{\pi}{2},0\right)\,,
\end{align} 
specifying the coordinates of the quark as a function of its proper time $\lambda$. In writing the above, we used spherical symmetry to place the quark in the black hole equatorial plane. In addition, any action we use will be invariant under reparameterisations of this world-line so we will choose the convenient gauge in which $\lambda = t$, i.e. we parameterise the world-line using the time coordinate of the ambient black hole spacetime. 

Second, we have to describe the flux tube and we do this by introducing the worldsheet
\begin{align}\label{eq:string_parametrisation}
    X^\mu(\tau, \sigma) = \left(t(\tau, \sigma), r(\tau, \sigma), \frac{\pi}{2}, 0\right)\,,
\end{align}
specifying the coordinates of the flux tube as a function of worldsheet coordinates $(\tau, \sigma)$. Again, we used spherical symmetry to place the flux tube in the equatorial plane. Since we are discussing a classical solution, we can pick a more convenient choice of worldsheet coordinates that simplify the problem. For the worldsheet time-like coordinate, we choose the time coordinate of the ambient spacetime so that $\tau = t$. For the space-like coordinate, we choose the radial coordinate of the ambient spacetime so that $\sigma = r$. This choice does not allow the string to move in the $(\theta, \phi)$ directions but this does not pose any problems. Indeed, in a static solution we expect the string to lie along a radial direction and we will verify below that this solves the equations of motion. Finally, we need to impose that the quark is located at the end of flux tube. We do this by demanding that the maximum $\sigma$ coordinate is $r(t)=r_q$ i.e. the equilibrium position of the quark. 

In order to derive the equations of motion of the quark and flux tube system in the black hole background, we start with the following action
\begin{align}
    S = -m_q \int_{\mathcal{C}} d\lambda + e \int_{\mathcal{C}} A_0 dt - \mathcal{T} \int_{\mathcal{S}} d^2\sigma \sqrt{-\gamma}\,,
\end{align}
where $\mathcal{C}$ denotes the quark worldline, $\mathcal{S}$ denotes the string worldsheet, $\mathcal{T} \approx \, \Lambda_N^2$ is the tension of the flux tube and $A_0$ is the black hole gauge field given in~\eqref{eq:EMfields}. Additionally, we use $\gamma$ to denote the induced metric on the worldsheet and it is given by
\begin{align}
    \gamma_{ab}= \frac{\pd X^\mu}{\pd \sigma^a}\frac{\pd X^\nu}{\pd \sigma^b} g_{\mu\nu} = \begin{pmatrix}
    -U(r) & 0 \\
    0 & U(r)^{-1}
    \end{pmatrix}.
\end{align}
In writing the induced metric, we used $U(\sigma)$ the blackening factor given in the RN metric in eq.~\eqref{eq:RNmetric} and the string parameterisation in~\eqref{eq:string_parametrisation}. Using this induced metric and our gauge choice, the action reduces to
\begin{align}
    S = -m_q \int dt \sqrt{U(r(t)) - U(r(t))^{-1} \dot{r}^2} - \frac{eQ_{\rm BH}}{4\pi} \int\frac{dt}{r(t)} - \mathcal{T} \int dt\; r(t)\,,
\end{align}
where we have performed the integral over the worldsheet coordinate $\sigma$. In line with our intuition, the system reduces to the single variable $r(t)$ which determines the quark position. The string simply stretches between the black hole horizon\footnote{The string may extend into the black hole and perhaps even get close to the singularity but this is irrelevant for the solution we present. Indeed, the end of the string that is at/behind the horizon contributes only a constant to the action above and therefore does not affect the equations of motion.} and the quark. 

We can now derive the equations of motion from the above action and determine a solution for $r(t)$. Our task is simplified by the fact that we are seeking a static solution for which $r(t) \equiv r_q$ is a constant. After we impose this on the equation of motion, we get
\begin{align}\label{eq:EOM}
    \frac{U'(r_q)}{\sqrt{U(r_q)}} = \frac{e Q_{\rm BH}}{2 \pi m_q r_q^2} - \frac{2 \mathcal{T}}{m_q}.
\end{align}
This is a complicated polynomial and it is difficult to find the general solutions in closed form. For this reason, we specialise to the simpler case of an extremal black hole for which:
\begin{align}
    U(r) = \left(1-\frac{GM_{\rm BH}}{r}\right)^2 = \left(1-\frac{Q_{\rm BH}}{4\sqrt{2}\pi \MPl r}\right)^2.
\end{align}
In this case the solution is:
\begin{align}\label{eq:r_eq_ex}
    r_q^{\rm (ex)} = \frac{1}{2\sqrt{\pi T}}\sqrt{\left(e - \frac{m_q}{\sqrt{2}\MPl}\right) Q_{\rm BH}}\,.
\end{align}
Note the appearance of the weak gravity conjecture (WGC) \cite{Arkani-Hamed:2006emk} under the square root. This is expected since we need the quark to be repelled from the black hole to be able to support a stretched string and this condition is precisely the weak gravity conjecture. In addition, requiring that the quark is outside the black hole gives
\begin{align}
    Q_{\rm BH} \leq \frac{2 \MPl^2}{\mathcal{T}}\left(e - \frac{m_q}{\sqrt{2}\MPl}\right) 
    \Longleftrightarrow M_{\rm BH} \leq \frac{2\sqrt{2} \MPl^3}{\mathcal{T}}\left(e - \frac{m_q}{\sqrt{2}\MPl}\right).
\end{align}
For Standard Model QCD, this corresponds to black holes lighter than about $10^{31} - 10^{32}$ grams.  Of course this number can take other values if we discuss a hidden QCD-like sector but it is remarkable that even for real-world QCD, we get black hole masses that encompass the interesting window for primordial black holes to be the dark matter~\cite{Carr:2009jm,Carr:2020xqk}. (We note however, that some of effects that we discuss do not occur in real-world QCD, since they will only be relevant in the large-$N$ limit.)

Another interesting limit where we can solve eq.~\eqref{eq:EOM} in closed form is the limit of massless quarks. In this case we obtain
\begin{align}\label{eq:r_eq_ex_massless}
    r_q^{(\text{massless $q$})} = \frac12 \sqrt{\frac{eQ_{\rm BH}}{\pi \mathcal{T}}}\,.
\end{align}
This is of course the limit of~\eqref{eq:r_eq_ex} as $m_q \rightarrow 0$. Recall, however, that we derived equation~\eqref{eq:r_eq_ex} for extremal black holes but no such assumption was made to arrive at equation~\eqref{eq:r_eq_ex_massless}. In this sense, as long as $eE>\Lambda_N^2$, all black holes look extremal to massless quarks. In this limit, to ensure that the quark is outside the black hole, we must require that
\begin{align}
    r_q^{(\text{massless $q$})} \geq r_+
\end{align}
which is not satisfied for $Q_{\rm BH} \rightarrow 0$ as expected. However, this inequality can always be satisfied for any $Q_{\rm BH} \neq 0$ if the flux tubes have low enough tension. Here it is important to note that black hole evolution is more difficult to analyse in the massless quark limit since the Schwinger rate in~\eqref{eq:Schwinger_rate_quarks} is potentially fast, and not under control. Nonetheless, we still expect quarks to be produced and confined at long distances so our qualitative picture still applies.

Finally, we comment on states composed of $k$ quarks that will in turn be attached to $k$ strings that extend to the black hole horizon. In the massless quark limit, we can see from~\eqref{eq:r_eq_ex_massless} that the larger tension $k\mathcal{T}$ compensates exactly for the larger charge $ke$ of the composite state. Corrections to the tension controlled by $1/N$~\cite{Athenodorou:2021qvs} can change this cancellation of $k$. In addition, the presence of quark masses can also prevent an exact cancellation between factors of $k$ in the charge and tension. Both of these effects are expected to cause the quark layer to thicken forming a quark `atmosphere' if bounds states are relevant. 

\subsubsection{The critical charge and corona formation}
In the previous section, we found the classical solution describing a flux tube emanating from a black hole and supported by the forces on a quark at its end. In addition, we saw that this solution only exists, in the sense that the string is outside the black hole horizon, for a subset of the full parameter space spanned by $\{e, M_{\rm BH}, Q_{\rm BH}, \mathcal{T}, m_q\}$. We now turn to a related question which is whether these strings can be spontaneously produced around the black hole. In section~\ref{sec:flat_space} we discussed Schwinger pair production in a capacitor, and saw that the formation of these flux tubes requires a region where the electric field satisfies $eE > \Lambda_N^2$. In this section we discuss the analogous condition in the context of black holes and argue that it is only satisfied for a subset of charged black hole solutions. We will identify this subset numerically for generic BH solutions but to get some intuition we will start by focusing on extremal black holes. 

Extremal black holes have a charge related to their mass by:
\begin{align}
    Q_{\rm BH} = \frac{M_{\rm BH}}{\sqrt{2}\MPl}.
\end{align}
For this value of the charge $r_\pm$ coincide and the inner and outer horizon are both at the same coordinate radius. The electric field at the horizon is maximal and given by
\begin{align}\label{eq:eBH_E-field}
    E = \frac{8\sqrt{2}\pi \MPl^3}{M_{\rm BH}} = \frac{8\pi \MPl^2}{Q_{\rm BH}}\,.
\end{align}
Surprisingly, smaller extremal black holes (with less electric charge) have larger electric fields near their horizon. It is then easy to see that the very large black holes cannot have an electric field which is strong enough to produce a black hole corona. Indeed, for an extremal BH, the critical charge and mass is found by imposing $e E = \Lambda_N^2$ and gives
\begin{align}
    \label{eq:critCharge}
    Q_{\rm crit} = 2\pi e\frac{ \MPl^2}{\Lambda_N^2} \quad ; \quad M_{\rm crit} = \sqrt{2} \MPl Q_{\rm crit},
\end{align}
with $e$ being the (dark) EM gauge coupling. This is the charge and mass of the largest extremal black hole that can produce quark--antiquark pairs and potentially form a corona. Any BH with charge (mass) larger than $Q_{\rm crit}$ ($M_{\rm crit}$) does not populate a quark corona, and just give us extremal BH with exponentially long lifetime. For charges (masses) below the critical values, the formation of the corona depends on the concrete values of $Q_{\rm crit}$ ($M_{\rm crit}$).

\begin{figure*}[t]
	\centering
	\includegraphics[width=0.6\textwidth]{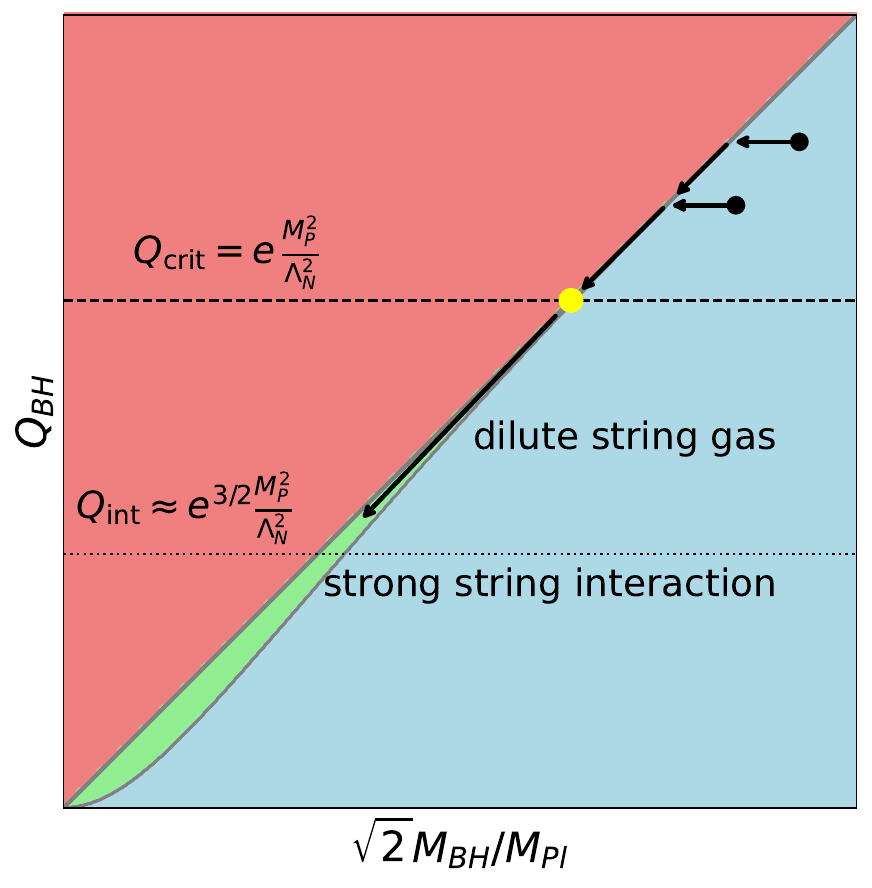}
		\caption{ \textbf{Qualitative evolution of large-$N$ black holes}. The shaded blue region corresponds to BH solutions without a corona due to their small electric field near the horizon, $eE< \Lambda_N^2$. The shaded green corresponds to BH solutions with electric field $eE>\Lambda_N^2$ for which the formation of a quark corona is possible. The line $Q_{\rm BH} = \sqrt{2}M_{\rm BH}/M_{\rm Pl}$ corresponds to BH extremality condition which separates the previous regions from the super-extremal region (in shaded red), forbidden by cosmic censorship. The black dots correspond to sub-extremal BH solutions with initial mass and charge such that $eE\ll \Lambda_N^2$. As these BH have $Q_{\rm BH}> Q_{\rm crit}$, we note that their evolution is such that $\dot{M}_{\rm BH}\gg \dot{Q}_{\rm BH}\approx 0$ and they always flow to extremality. After reaching extremality, they track the extremal line and reach the yellow point, $(M_{\rm crit},Q_{\rm crit})$, which becomes an attractor (see section \ref{sec:dynamics}). After this, the BHs populate a quark corona and evolve near extremality until the charge in the corona becomes $Q_{\rm cor}=\frac{1}{e}\frac{Q_{\rm BH}^2}{Q_{\rm crit}}$. As the baryon emission is exponentially suppressed, we can solve for the BH charge and obtain the charge below which string-string interactions become important, $Q_{\rm int}\approx e^{3/2}\frac{\MPl^2}{\Lambda_N^2}$. Below this charge our qualitative description of a near-extremal BH breaks down.}
	\label{fig:BH_corona-parameter_space}
\end{figure*}

\subsubsection{Solving Einstein equations}
\label{sec:Einstein_eq_sol}
For BHs with $Q_{\rm BH}\lesssim Q_{\rm crit}$, the electric field can satisfy $eE\gtrsim \Lambda_N^2$ and the formation of the quark corona occurs through Schwinger pair production. This rate is exponentially suppressed and therefore the process is very slow in the case of heavy quarks, $m_q^2\gtrsim (eE-\Lambda_N^2)$. In this section we find a solution for such spacetime with many strings emanating radially from a charged BH. Their endpoints contain quarks, and this configuration is sustained by a balance between gravity, tension, and electromagnetic repulsion. It is not a bad approximation to consider a configuration with multiple strings emanating from the BH as a quasi-static configuration.  We solve the Einstein equations and obtain the metric for the region between the BH horizon and the quark corona, located at $r_{\rm cor}$. For $r\gg r_{\rm cor}$ we match this solution to a standard RN solution. 

As we are looking for the solution in the region $r\lesssim r_{\rm cor}$ we can neglect the contribution to the energy-momentum tensor from the quarks at the endpoints of the strings by using a standard Gauss law argument. The relevant part of the energy-momentum tensor is
\begin{equation}\label{eq:tot_E-M_tensor}
    T_{\rm tot}=T_{\rm EM}+T_{\rm strings}\,.
\end{equation}
Due to spherical symmetry, the energy-momentum tensor for radially oriented strings reads
\begin{equation}\label{eq:EnergyMomentum_tensor}
    \tensor{(T_{\rm tot})}{^\mu_\nu}=-\text{diag}(\rho (r),\rho (r),0,0)\,, \text{ with: }\rho(r)=\frac{n_{str} \Lambda_N^2}{4\pi r^2}\,,
\end{equation}
that is, both the energy density and pressure are given in terms of the string tension, which is radially oriented. 

Thanks to spherical symmetry one can easily solve the Einstein equations. Since the configuration is static~\cite{Misner:1973prb}, the only equation to solve is $m^\prime(t,r)=4\pi Gr^2 (-T_{\rm tot})^t_t$, where $T_{\rm tot}$ is the total energy-momentum tensor in Eq.~\eqref{eq:tot_E-M_tensor} and the prime indicates derivative with respect to $r$. Solving this equation  one obtains the metric for regions inside the spherical shell given by the quark corona,
\begin{align}
\label{eq:BHCoronaMetric}
   ds^2=-f(r)dt^2+f(r)^{-1}dr^2+r^2d\Omega \,,\nonumber\\
   f(r)=1-\frac{2G m_0}{r}+\frac{G Q_{\rm in}^2}{4\pi r^2}-2 G  n_{\rm str}\Lambda_N^2 \,.
\end{align}
Here $Q_{\rm in}$ is the electric charge of the BH inside the corona, $m_0$ the mass of the BH, and $n_{\rm str}$ the number of strings emanating from the BH (radially oriented in a spherically symmetric way). Both, the mass and charge in the corona, will not be relevant to for the energy-momentum tensor due to the usual Gauss law argument. Therefore, this solution is valid for 
\begin{equation}
    R_{\rm BH}\lesssim r < r_{\rm cor}\,.
\end{equation}

The new term, $2Gn_{\rm str}\Lambda_N^2$, is important since it tells us that horizon will be displaced. The new horizon is obtained by solving for $f(\tilde{r}_\pm)=0$, which gives
\begin{equation}\label{eq:moved_horizon}
    \tilde{r}_\pm = \frac{Gm_0 \pm \sqrt{(Gm_0)^2 - GQ_{\rm in}^2(1-2n_{\rm str}G\Lambda_N^2)/4\pi }}{1 - 2 n_{\rm str}G\Lambda_N^2}\,.
\end{equation}
As expected, the corrections to the position of the horizon are of order $\sim n_{\rm str}G\Lambda_N^2$. Note also that the $r\rightarrow \infty$ limit of $f(r)$, namely $(1-2n_{\rm str}G\Lambda_N^2)$, appears in the expression for $\tilde{r}$ and is the only modification compared to the RN case. If the above metric were valid for all $r$, this $r$-independent factor means that the space is not asymptotically flat. Of course, this is obvious given that the energy density in~\eqref{eq:EnergyMomentum_tensor} does not vanish fast enough for flat asymptotics. However, one should be cautious with this conclusion since the smeared approximation we used to derive~\eqref{eq:BHCoronaMetric} fails very far away from the BH, as the strings become more and more separated from each other. In any case, for our application, we will only use the solution presented above between the BH horizon and the layer of quarks. Finally, we note that at $2n_{\rm str}G\Lambda_N^2\sim 1$, the solution is expected to break down.

At distances larger than the size of the corona, $r\gg r_{\rm cor}$, the solution is matched to the standard RN solution with charge $Q_{\rm tot}=Q_{\rm in}+Q_{\rm cor}$ and mass $M_{\rm tot}=m_0+m_{\rm cor}$. The metric for $r\gg R_{\rm BH}+r_{\rm cor}$ is now given by

\begin{align}
\label{eq:RNmetric_outside}
ds^2 &= -U(r) dt^2 + U(r)^{-1} dr^2 + r^2 d\Omega^2\,,\\&
U(r) = 1 - \frac{2GM_{\rm tot}}{r} +  \frac{G Q_{\rm tot}^2}{4\pi r^2}\,.
\end{align}

While the metric solutions above describe consistent BH backgrounds, the formation of these objects is easiest to understand by starting with charged BHs without a corona and then producing the corona dynamically. We expect this to happen (eventually) for charged black holes whether their charge is above or below the critical value~\eqref{eq:critCharge}. If one starts with a black hole whose charge is below~\eqref{eq:critCharge}, the evolution is more complicated and we leave its description for future work. In the following section, we will study corona formation in the simpler case where the initial BH charge is above~\eqref{eq:critCharge}.

\section{Dynamics}\label{sec:dynamics}
Any BH in a $SU(N) \times U(1)$ theory with matter transforming as $q\sim (\mathbf{N},1)$ will evolve towards extremality. The mechanism is similar in spirit to the one presented in \cite{Hiscock:1990ex} (see also \cite{Bai:2019zcd}) where it is argued that near-extremal black holes can approach extremality if the only charged particles in the spectrum are more massive than the black hole temperature. In our case, however, the suppression is two-fold. The first suppression factor is the same thermal factor although this may be small (i.e. unsuppressed) when the black hole is away from extremality. The second factor is the combinatoric suppression in the large-$N$ limit. This is present because the theory confines at long distances and the only asymptotic states are colour singlets, i.e. mesons or large-$N$ baryons. With this restriction in mind, the black hole may radiate (i) neutral particles such as mesons, photons and gravitons which only reduce the black hole mass or (ii) a charged baryon; which is made up of $N$ quarks leading to a suppression of this process by the combinatoric $e^{-N\log N}$ factor. As such, there is no process that allows the black hole to lose charge efficiently\footnote{Of course, if the charged black hole exists in a plasma like in the early universe, it may accrete oppositely charged particles and reduce its charge. In this discussion we are considering black holes in a vacuum asymptotically Minkowski space.}\footnote{Similarly, superrandiance of baryons will be exponentially suppressed. We expect similar suppression factors for other extended objects.} and it is driven towards extremality. This result is independent of the radiation mechanism, particle masses, and confinement scale. Recent studies including quantum effects in Maxwell theory coupled to gravity have shown that a BH with SM electric charge is only exponentially close to extremality~\cite{Brown:2024ajk}. For the purpose of this paper, slowing down the approach to extremality strengthens our results. That said, we do not include these effects and assume that an extremal BH with zero temperature can be reached. In this sense, our argument is conservative and we leave the inclusion of quantum corrections for future work.

Once these black holes are close to extremal their evolution stops for exponentially long times but will decay eventually since they are not protected by any symmetry. Their stability is ensured by cosmic censorship and the combinatoric suppression relevant when radiating baryons. Upon reaching extremality via the processes discussed in the previous paragraph, these black holes can no longer radiate neutral particles. This can be seen by inspecting eq.~\eqref{eq:RNtemperature} which shows that the black hole temperature vanishes in the limit $M_{\rm BH}\rightarrow 2 \sqrt{2} \MPl Q_{\rm BH}$. Said differently, when the black hole has $M_{\rm BH}=2\sqrt{2}\MPl Q_{\rm BH}$, losing more mass would lead to a super-extremal black hole. These have naked singularities and violate cosmic censorship which is believed to be true in general relativity~\cite{Wald:1997wa}. The only way for these black holes to continue to lose mass is by emitting a Weak Gravity particle that has $2\sqrt{2} e q \MPl > m$ \cite{Arkani-Hamed:2006emk}. The only candidate in the theory we consider is the baryon but we already saw that this process is suppressed by combinatorics. This means that the black hole lifetime would be extended by a factor that grows exponentially with the number of colours $N$, $\tau_{\rm BH}\propto e^{N\log N}$.

At times much longer than $\tau_{\rm BH}$ the BH always develops a quark corona.  There is a color flux tube (confining string) between the quark, suspended outside the BH, and the antiquark that falls behind the horizon.  As the BH populates a denser quark corona, it loses charge and gets smaller, following closely the $M_{\rm BH}=2\sqrt{2}\MPl Q_{\rm BH}$ line. Interestingly, since the BH always stays extremal or near-extremal, the electric field increases as $E\sim \frac{\MPl^2}{Q_{BH}}$ instead of decreasing! This leads to an acceleration of the formation of the quark corona, an interesting difference with respect to the case of the capacitor in section~\ref{sec:flat_space}. Numerical studies necessary to properly understand the evolution in a more quantitative way are beyond the scope of this work and will be presented elsewhere. 

\subsection{An example with $Q_{\rm initial}>Q_{\rm crit}$}

As an example, let us take a (non-extremal) BH with $Q_{\rm BH}>Q_{\rm crit}$ and study qualitatively its evolution (see Fig.~\ref{fig:BH_corona-parameter_space} for more details). For this BH, the Schwinger pair production of quark-antiquarks that leads to a quark corona is not possible initially as the electric field at the surface is not large enough, $eE\lesssim \Lambda_N^2$. Neutral Hawking radiation will be emitted as we have $T_H\neq 0$. Radiation of mesons, (hidden) photons and gravitons  will lead to loss of BH mass without $U(1)$ charge emission, since the production of baryons is very suppressed. In the $(M_{\rm BH},Q_{\rm BH})$ plane the BH initially moves horizontally approaching the extremal bound. 
Interestingly, an external visible sector observer would see a BH that is evaporating through neutral Hawking radiation (possible including SM neutral particles) and losing temperature instead of increasing it! This is due to the fact that the BH is approaching extremality of the dark $U(1)$ charge, which is not measurable by the SM observer.

As discussed previously, once extremality is reached the only allowed process that can induce BH discharge, or evaporation of any kind, is the spontaneous emission of a charged baryon (lightest charged, color neutral state) due to the Schwinger effect. This process is exponentially suppressed (see sec.~\ref{sec:flat_space}) as
\begin{equation}\label{eq:rate_BH_baryon}
    \Gamma \sim \left ( eE \right )^{1/2} e^{-S_{\rm Sch.}}e^{-N\log N}\,,
\end{equation}
where the first exponential is the usual Schwinger-type suppression,
\begin{equation}
    S_{\rm Sch.}=\frac{M_{\rm B}^2}{q_{\rm B}}\frac{1}{eE}\,.
\end{equation}
Given that $M_{\rm B}\propto N$ and the quantised baryon charge is $q_{\rm B}=N$, for $eE\sim \Lambda_N^2$, then as soon as we start forming the corona we have
\begin{equation}
    S_{\rm Sch.}\gtrsim N\,.
\end{equation} 
Note that the electric field will increase as the BH evolves, meaning that the Schwinger rate tends to increase. The second exponential factor in Eq.~\eqref{eq:rate_BH_baryon}, $e^{-N\log N}$, corresponds to a \textit{combinatoric} suppression that arises due to the formation of baryons from nucleating $N$ quarks with the right colours and is constant in time. 

Successive emission of baryons will lead to the discharge of the BH tracking the extremal line -- this occurs, intuitively, because the BH is (slowly) losing charge through the emission of baryons followed by neutral Hawking radiation which drives the resulting BH back to extremality. For a BH with initial charge $Q_{\rm BH}\gg Q_{\rm crit}$, this process will be very slow due to Eq.~\eqref{eq:eBH_E-field} and the extremal BH will only reach the critical charge, $Q_{\rm BH}=Q_{\rm crit}$, after an exponentially long time,
\begin{equation}
    \tau \gg \left ( eE \right )^{-1/2} e^{N\log N}\,.
\end{equation}
After this, $U(1)$ charges below the critical value $Q_{\rm BH}\lesssim Q_{\rm crit}$ will be reached and the corona will form. 

The discussion above is just a qualitative description that justifies that every BH will eventually form a quark corona although, possibly, an exponentially long time after becoming extremal. In the next section we study a quasi-static configuration where the BH is attached to many strings radially oriented.

\subsection{Corona production}
As the large black holes described in the previous section lose charge, they will eventually cross into the $Q_{\rm BH} < Q_{\rm crit}$ where the production of flux tubes attached to quarks can happen. In this regime, as we will explain below, we do not have complete control over the equations describing the evolution of the black hole, except for the initial stages. For this reason, we will only describe this initial evolution leaving studies of the full black hole and corona system for future work. 

We can describe the evolution \`{a} la Hiscock and Weems~\cite{Hiscock:1990ex}.  As we cross below $Q_{\rm crit}$ the system of equations needs to be modified so it describes three quantities: $Q_{\rm BH},\, Q_{\rm cor},\, M_{\rm BH}$. These are sufficient to determine the black hole solution with smeared strings described in section~\ref{sec:Einstein_eq_sol}.

In the initial stages of evolution, we can still trust the Schwinger pair-production picture and write equations analogous to those in~\cite{Hiscock:1990ex} describing the temporal dependence of $\{Q_{\rm BH}, Q_{\rm cor}, M_{\rm BH}\}$. In the large-$N$ limit, and moving down the extremality line, the time-variation of $\{Q_{\rm BH}, Q_{\rm cor}, M_{\rm BH}\}$ are all slow. First, the baryon emission rate is still exponentially suppressed in $N$ and may be safely ignored. We therefore have $\dot{Q}_{\rm BH} = -\dot{Q}_{\rm cor}$, by charge conservation, and both of these are exponentially small because the bounce action in~\eqref{eq:Schwinger_rate_quarks} diverges. All that remains is the emission rate of neutral particles (i.e. photons, gravitons, etc.) in $\dot{M}_{\rm BH}$ and this is suppressed by powers of the black hole temperature which vanishes near on the extremality line. 

In this regime, the black hole will grow a corona of quarks tethered by flux tubes to its horizon. The qualitative picture is given by the following considerations. Imagine starting with an extremal BH, whose electric field produces pairs of $\bar{q}q$ (see~\cite{Lin:2024jug} for finite size effects, which may be important when the Compton wavelength of the charged particle is large.). The antiquark falls into the black hole and the quark is pushed away by the electric field to its equilibrium position \eqref{eq:r_eq_ex}. This process reduces the black hole charge and moves the horizon outwards. The position of the horizon is initially given by the RN horizon with reduced $Q$, up to corrections due to the energy density in the corona which is controlled by $G \Lambda_N^2$, see Eq.~\eqref{eq:moved_horizon}. As the quark emission continues, and the number of strings in the corona increases, the spacetime will depart from the RN geometry and is described instead by the metric in~\eqref{eq:BHCoronaMetric}. This happens when the number of strings is $n_{\rm str.} \gg 1$ so that the system has approximate spherical symmetry. As the number of strings increases further, one of two things can happen that invalidate our description of the spacetime. First, the metric in~\eqref{eq:BHCoronaMetric} becomes singular for $n_{\rm str.} > \MPl^2/\Lambda_N^2$ and we cannot trust this picture in that regime. Additionally, our assumption of non-interacting strings breaks down when the number of strings per $\Lambda_N^{-2}$ area becomes of order $N$. This is the bound discussed in section \eqref{sec:dilute_string_gas} and is summarized in equation~\ref{eq:QcorBound}. 

\subsection{The dilute string gas approximation}
\label{sec:dilute_string_gas}
Throughout the discussion of the quark corona we have neglected string-string and (Coulomb-like) quark-quark interactions. In this section we discuss the validity of this approximation. In the large-$N$ limit, strings become non-interacting. The same occurs with the Coulomb repulsion between quarks once the $U(1)$ gauge coupling has been rescaled as $e\sim 1/\sqrt{N}$. However, the collective effect of a large number of these objects may cause the breakdown of this approximation. This will occur whenever there are $\mathcal{O}(N)$  strings crossing a section with an area of order $\Lambda_N^{-2}$, or $N$ quarks in a volume $\sim \Lambda_N^{-3}$. 

Let us for simplicity consider an extremal BH. The area is given in terms of the total charge by $A\sim \frac{Q_{\rm BH}^2}{M_{\rm Pl}^2}$. This means that in order to have less than $\mathcal{O}(N)$ strings for each patch of area $\Lambda_N^{-2}$, the following condition has to be satisfied
\begin{equation}
\label{eq:QcorBound}
    Q_{\rm cor}\lesssim N\frac{\Lambda_N^2}{M_{\rm Pl}^2} Q_{\rm BH}^2\,.
\end{equation}
Recalling that we scale the gauge coupling $e \sim 1/\sqrt{N}$ in the large-$N$ limit, the above quantity is more conveniently written as
\begin{align}
    \label{eq:QcorBound2}
    Q_{\rm cor}\lesssim \frac{1}{e}\frac{\QBH^2}{Q_{\rm crit}}.
\end{align}
As explained above, for $\QBH \lesssim Q_{\rm crit}$, we expect the black hole to quickly produce a corona. In this regime, we can neglect the slow baryon emission and approximate the total charge as $Q_{\rm cor} +\QBH =Q_{\rm tot} \approx Q_{\rm crit}$. Using this approximation, we can solve~\eqref{eq:QcorBound2} for $Q_{\rm int}$, the BH charge at which the string interactions cannot be ignored. This gives
\begin{align}
    Q_{\rm int} \approx e^{3/2}\frac{\MPl^2}{\Lambda_N^2}
\end{align}
where we wrote the result in the large-$N$ limit. Note that this charge is parametrically smaller than $Q_{\rm crit}$, by a factor of $\sqrt{e}\sim N^{-1/4}$. As such, there is a large controlled regime where we can trust our solution before string-string interactions become important. 

As the BH evaporates and becomes smaller the electric field grows, and the corona becomes denser and larger (that is, longer strings). Indeed, at some point the length of the strings could be larger than the BH size. However, reaching this limit depends on other parameters of the theory such as the quark mass. This occurs because, as described in Sec.~\ref{sec:dynamics}, the Schwinger action (electric field) decreases (increases) with time. When $S_{\rm Sch} \sim \frac{m_q^2}{eE}\sim \frac{m_q^2}{N^{1/4}\Lambda_N^2} = \mathcal{O}(1)$, quark-antiquark pair production becomes  fast and our qualitative description breaks down. For this reason, our present discussion is only valid until the Schwinger action is $S_{\rm Sch}\sim \mathcal{O}(1)$ or the BH charge decreases below $Q_{\rm int} \sim e^{3/2}\frac{\MPl^2}{\Lambda_N^2}$, whichever occurs first.

\section{Long-lived extremal BHs and the swampland}\label{sec:swamp}
The stability of the extremal BHs in the limit of large number of colours may raise questions about the compatibility of the EFT with the weak gravity conjecture (WGC)~\cite{Arkani-Hamed:2006emk} and other swampland conjectures~\cite{Vafa:2005ui,Brennan:2017rbf,Palti:2019pca,vanBeest:2021lhn}.
The WGC demands the existence of super-extremal particles that can, in particular, allow extremal BHs to decay. 
For standard $U(1)$ gauge fields this means that there must exist a charged particle in the spectrum with a mass satisfying $m<eq M_{\rm Pl}$, where $e$ is the gauge coupling. 

Given an extremal BH solution with charge $Q_{\rm BH}$, the WGC can be used to derive lower bound for the Schwinger emission rate. As we will now show, the BH solutions above in Sec.~\ref{sec:including_gravity} violate this bound exponentially. Let us estimate the Schwinger emission rate in the weak field limit. This process is exponentially suppressed by the Schwinger action,
\begin{equation}
    S_{\rm Sch.}=\frac{m^2}{qeE}\lesssim qeQ_{\rm BH}\,,
\end{equation}
where we have used both the WGC bound and the electric field for an extremal BH, $E=\frac{M_P^2}{Q_{\rm BH}}$. Therefore, for a given extremal BH, the WGC predicts
\begin{equation}
    \Gamma \gtrsim M_{\rm Pl} \left (\frac{e}{Q_{\rm BH}}\right )^{1/2}e^{-qeQ_{\rm BH}}\,.
\end{equation}

The large-$N$ theories we studied above, however, have an additional exponential suppression factor (see Eq.~\eqref{eq:rate_BH_baryon}). For a baryon of the same charge ($qe$) as the one in the previous equation, the emission rate is
\begin{equation}
    \Gamma \gtrsim M_{\rm Pl} \left (\frac{e}{Q_{\rm BH}}\right )^{1/2} e^{-qeQ_{\rm BH}}\,e^{-N\log N} \,,
\end{equation}
so we violate the rate implied the WGC exponentially. This exponential factor, $e^{-N\log N}$, is precisely what allows the BH to remain in a meta-stable state for exponentially long times before continuing its evolution (e.g. when $Q>Q_{\rm crit}$). As we discussed above, this arises due to the fact that in the confined phase of the theory, the only charged degrees of freedom are baryons. These objects are not point-like particles but a composite solitonic objects with $M_{\rm B}R_{\rm B}\sim N$. Similar violations are expected, for example, in theories with monopoles confined by strings. 

In this section we will study the theoretical constraints imposed on the number of colours by the species bound~\cite{Veneziano:2001ah,Dvali:2007hz,Dvali:2007wp}, which turns out to be the most constraining requirement, as well as completeness of the spectrum~\cite{Heidenreich:2021xpr}. As before, we consider theories based on $SU(N)\times U(1)$ with quarks in transforming as $q\sim (\mathbf{N},1)$. 

\subsection{Species bound and weak gravity conjecture in the large-$N$ limit}
Consistent QFTs have an upper bound for the number of species -- that is, the number of degrees of freedom below a certain scale -- given by the relation \cite{Dvali:2007hz,Dvali:2007wp}
\begin{equation}\label{eq:species_bound}
    \MPl^2\geq  N_{\rm dof}\Lambda_{\rm UV}^2\,,
\end{equation}
where $N_{\rm dof}$ is the number of species below the cut-off, $\Lambda_{\rm UV}$~\footnote{We note that these 4d species bounds are analogous to the definition of the 4d Planck scale in UV complete theories including extra-dimensional theories or superstring theory.}. For the large-$N$ theories that we consider, it is natural to expect that this coincides with the number of gluons, scaling as $N_{\rm dof}\sim N^2$.
The effective field theory description will therefore be valid up to $\Lambda_{\rm UV}$. Above this scale, quantum gravity effects are expected to be important and our description will be inaccurate. For this reason we will refer to it as the quantum gravity or string scale, $\Lambda_{\rm UV}\sim l_s^{-1}$. 

For the theory to be consistent, both the quark mass and the confinement scale must lie below the UV cutoff, $m_q\,,\Lambda_N < \Lambda_{\rm UV}$.  This implies that the baryon size (roughly given by $\Lambda_N^{-1}$) is above the quantum gravity length scale and, crucially, that its mass is below the 4d Planck scale 
\begin{equation}\label{eq:species_bound2}
    M_{\rm B}\sim N(m_q+ \Lambda_N)< N\Lambda_{\rm UV}\lesssim \MPl\,\rightarrow N\lesssim \frac{\MPl}{\Lambda_{\rm UV}}.
\end{equation}
Similar bounds have been obtained in \cite{Kaplan:2019soo,Kaplan:2020tdz} using different arguments, including unitarity and causality.
Interestingly, we find that the species bound~\eqref{eq:species_bound} automatically implies the WGC~\cite{Arkani-Hamed:2006emk}, which   requires the baryon to be a super-extremal particle,
\begin{equation}\label{eq:WGC}
    M_{\rm B}<\MPl\rightarrow M_{\rm B}\ll eq_{\rm B}\MPl=\sqrt{N}\MPl\,.
\end{equation}
In this class of theories, the WGC bound is less restrictive than the species bound. Furthermore, we note that satisfying this inequality implies that the quark satisfies the WGC as well.

As a last comment we note that both, the species bound and the WGC, are in conflict with some of the results in \cite{Profumo:2025var}. In this reference, the baryon-BH transition requires that the baryon size is below its Schwarschild radius, $R_{\rm B}\lesssim \frac{M_{\rm B}}{\MPl^2}$, therefore requiring a number of colours which is $N\gtrsim \frac{\MPl^2}{\Lambda_N^2}$, in conflict with~\eqref{eq:species_bound2} and~\eqref{eq:WGC}. 
In our scenario, we do not require the baryon size to be below its Schwarschild radius. Instead, the stability of the BHs that we studied comes from the (gauge) $U(1)$ charge and the difficulty in emitting charged baryons in the large-$N$ limit. Introducing an additional $U(1)$ is also crucial to for the process of quark deconfinement by large electric fields and, in the context of BHs, for the formation of a quark corona.

\subsection{Completeness of the spectrum}
\label{sec:completeness}
To ensure the absence of global symmetries, a consistent QFT coupled to gravity must contain all the charges compatible with Dirac quantisation,
\begin{equation}\label{eq:Dirac_quantisation}
    q_eq_g=n/2\,.
\end{equation}
That is, we must have particle (or multiparticle) states transforming in all the representations of the gauge group \cite{Heidenreich:2021xpr}. In the case of interest, we have a $SU(N)\times U(1)$ gauge theory coupled to gravity with quarks transforming as $q\sim (\mathbf{N},1)$. The quantised electric charge of baryons  $Q_{\rm B}=N$ naively suggests the presence of an electric $Z_N$ 1-form symmetry. 
However, before drawing such conclusion, the spectrum of magnetic monopoles has to be specified. 

In the theory of interest, the spectrum of monopoles that satisfies Dirac quantisation is given by two kinds of monopoles \cite{Corrigan:1976wk}. First, we have a standard monopole with $q_g=1$, and no chromomagnetic charge. Because of the YM sector, there is a second, qualitatively different kind of monopole -- a monopole with minimal magnetic charge, $q_{g_0}=1/N$, and chromomagnetic charge so that the state is consistent with Dirac quantisation.
This is analogous to the standard GUT monopole case with fractionally charged quarks. Since we satisfy Dirac quantisation for all integers we have a complete spectrum. 
For compact abelian groups this implies that there is no global electric 1-form symmetry.

\section{Phenomenology of large-$N$ black holes}
\label{sec:pheno}

The large-$N$ BH solutions presented above have interesting applications for phenomenology. Let us assume that the gauge group  $SU(N)\times U(1)$ is a hidden sector so that the BH only interacts with the SM sector gravitationally~\footnote{The interesting possibility where the dark $U(1)$ has kinetic mixing with electromagnetism will be considered elsewhere.}. When the large-$N$ BH approaches extremality, its temperature (measured at infinity) drops to zero, $T\rightarrow 0$. At this point, for $Q>Q_{\rm crit}$, the BH can only discharge through baryon emission, which as we have seen in previous sections is an exponentially suppressed process. This leads to long-lived extremal BHs that may survive until today contributing to the DM abundance. If the electric field satisfies $eE> \Lambda_N^2$, the formation of a quark corona around the BH leads to a new type of exotic, compact objects with potentially interesting implications for BH mergers and their associated gravitational wave signal. In this section we briefly discuss the former, leaving the latter for future work.

\subsection{\textit{Ultralight} primordial BH dark matter}
Primordial BHs are interesting dark matter (DM) candidates spanning many orders of magnitude in mass (see~\cite{Carr:2020xqk} for a review). The common lore is that in order to avoid fast Hawking evaporation and survive until today as a sizeable fraction of DM, PBHs must have a relatively large mass. 
Their lifetime is roughly given by \cite{Carr:2009jm}
\begin{equation}\label{eq:lifetime_PBH}
    \tau_{\rm PBH} (M_{\rm BH})\sim G^2M_{\rm BH}^3\,,
\end{equation}
indicating that PBH DM requires $M_{\rm BH}\gg 10^{17}$ grams. 

Remarkably, in $SU(N)\times U(1)$ theories, the charged BHs we have proposed are naturally long-lived 
\begin{equation}
    \tau\gtrsim e^{N\log N}\tau_{\rm PBH}(M_{\rm BH})\,,
\end{equation}
suggesting large-$N$ theories as a compelling stabilisation mechanism for light PBHs.  As an illustrative example, let us consider a BH of mass $M_{\rm BH}=1$ gram. Without the large-$N$ effects studied here, this BH would have a very large temperature of order of the GUT scale and evaporate in a short time, $\tau_{\rm PBH}\sim 10^{-28}$ seconds. Strikingly, in a $SU(N)\times U(1)$ theory a PBH with $U(1)$ charge and mass $M_{\rm BH}\sim 1$ gram can be cosmologically stable for $N\gtrsim 30$. This statement relies only on combinatorics and is independent of the radiation mechanism from the black hole (be it Hawking radiation or Schwinger pair production). In particular, we do not have to make assumptions about the particle masses or other scales in the problem (e.g. confinement scale $\Lambda_{N}$) as discussed earlier.

Such a large number of colours might seem too large to be realised in nature. Now we show that there is an interesting way to generate it effectively through a \textit{clockwork-like} theory~\cite{Giudice:2016yja}. Let us consider a theory with multiple YM sectors of the form
\begin{equation}
    SU(N_1)\times ... \times SU(N_n)\times U(1)\,.
\end{equation}
Our findings in previous sections can be easily generalised to this class of theories, which as we will show can ensure BH stability without requiring large number of colours. Indeed, with the right quark charge assignment, the theory resembles a theory based on the group
\begin{equation}
    SU(N_{eff})\times U(1)\,,\,\text{with: }\, N_{eff}=\Pi_iN_i\,.
\end{equation}

Let us illustrate this with a toy model based on
\begin{equation}
    SU(N_1)\times SU(N_2)\times U(1)\,,
\end{equation}
and quarks transforming as the bi-fundamental representation, $q\sim (\mathbf{N}_1,\mathbf{N}_2,1)$. In this case, when $N_1,N_2$ are co-prime numbers there is no $Z_{gcd(N_1,N_2)}$ 1-form symmetry and the spectrum is complete. 
Regarding the $U(1)$ gauge group, completeness requires the minimal magnetic charge to be $g_0=\frac{1}{N_1N_2}$. As in the $SU(N)\times U(1)$ monopole, $g_0$ also carries chromomagnetic charges (see \ref{sec:completeness}). Therefore, in this case, there is no electric 1-form symmetry for the $U(1)$ sector and the spectrum is complete.
 
Consider now a BH with $Q_{\rm BH} ,M_{\rm BH}$ above the critical values so that the electric field at the near-horizon region is $eE\lesssim \Lambda_N^2$. In this case, the life-time of the BH is at least
\begin{equation}
    \tau_{N_1N_2}\gtrsim e^{N_1N_2\log N_1N_2}\tau_{\rm PBH}(M_{\rm BH})\,.
\end{equation}
As an example, in a theory with $N_1=5$, $N_2=7$ (or similar co-primes) the life-time is longer than the age of the universe even for BHs as light as a $M_{\rm BH}\sim 0.01$ grams!
This toy model is far form being a complete theory of ultralight PBH dark matter, but illustrates how two dark sectors may conspire to stabilise the light PBHs for cosmological time scales, effectively generating the same stabilisation mechanism as a large $N$ theory, in the spirit of clockwork theories. It would be nice to explore concrete formation mechanisms for this kind of PBH that allow to produce them in the early universe in a way compatible with the DM abundance.

A final point that deserves mention is that the extremal primordial BH proposed here may be more easily produced in a universe with a net hidden baryon number. We note however, that such universe would have also a net $U(1)$ gauge charge. It is unclear to the authors whether this poses theoretical problems. Similar situations appear in \cite{Kaplan:2023fbl,DelGrosso:2025ygg}. It would be nice to find more explicit connections with the scenarios studied in these references.

\section{Conclusion}\label{sec:conclusion}
We have shown that large-$N$ $SU(N)\times U(1)$ theories coupled to gravity with quarks transforming as $q\sim (\mathbf{N},1)$ have a qualitatively new kind of BH solution with interesting implications. 
Due to confinement, the only colour singlet asymptotic states with electric charge are baryons -- states composed of $N$ quarks with totally antisymmetric contraction of their colour charge. We have shown that, in the large-$N$ limit, the emission of baryons due to Hawking radiation or via Schwinger pair production is exponentially suppressed by a factor scaling as $\sim e^{-N\log N}$ while Hawking radiation of neutral states is not suppressed by such factors. This implies that, above the critical charge $Q_{\rm BH} \gtrsim Q_{\rm crit} = 2\pi e\frac{M_{\rm Pl}^2}{\Lambda_N^2}$, any charged BH will evolve towards extremality regardless of their initial conditions. BHs in these theories are therefore much colder and long-lived than naively expected. An interpretation of this phenomenon is as follows. While the UV theory above the confinement scale is composed of elementary coloured particles with $U(1)$ charge (quarks), below confinement, the only charged states are baryons. As observed long ago by Witten~\cite{Witten:1979kh}, in the limit of a large number of colours, baryons are \textit{fluffy} solitonic objects with the product of their mass and size scaling as $M_{\rm B}R_{\rm B}\sim N$. This implies that the spontaneous pair production of baryons requires large fluctuations. These fluctuations are exponentially suppressed by a large instanton-like action, $S\gtrsim N\log N$. 

After an exponentially long time, $t\gg \tau_{\rm BH}\sim (eE)^{-1/2}e^{N\log N}$, and once the electric field at the horizon is large enough, $eE>\Lambda_N^2$, the (near) extremal BH will enter into a novel, exotic phase developing a \textit{quark corona} -- a cloud of deconfined quarks around the horizon. These quarks are deconfined because they are connected to the horizon by a colour flux tube that extends throughout a distance which can be parametrically larger than the confinement scale, $l\gg \Lambda_N^{-1}$. Interestingly, we have shown that energy conservation and cosmic censorship forbids the breaking of the string, independently of their length and the quark mass.  This quark corona is, in some sense, the electric analog of the magnetic monopole corona studied by Maldacena \cite{Maldacena:2020skw}. When the Schwinger rate  of quark--antiquark pairs is sufficiently suppressed -- that is, in the limit that BH evolution is very slow -- it is possible to solve Einstein equations and obtain a metric describing the space-time of a BH surrounded by a corona. The new horizon is displaced with respect to the standard Reissner-Nordstr\"om solution by a factor scaling as $\sim n_{\rm str}G\Lambda_N^2$.

This new kind of BH that we have found has potentially interesting  applications for phenomenology. As a particular example, we studied the case of \textit{ultralight} primordial BH dark matter. We  show that in these theories BH are exponentially long-lived even if the quark mass $m_q$ is much smaller than the Planck scale or the confinement scale, $\Lambda_N$. Indeed, in some regions of parameter space the BH is  exponentially long-lived even if the quark is massless. This occurs because once extremality is reached, the only way to discharge the BH is by Schwinger pair production of baryons. This process is doubly suppressed -- in addition to the usual Schwinger suppression by the baryon mass, $\Gamma \sim (eE)^{1/2} e^{-\frac{M_{\rm B}^2}{eq_{\rm B}E}}$, the process of baryon emission is $e^{-N\log N}$ suppressed. 

The mechanism that we studied can be easily generalised for stabilising light PBH in the spirit of a \textit{non-abelian clockwork} theory. Indeed, for theories of the type $SU(N_1)\times SU(N_2)\times ... \times U(1)$, the lifetime  grows as $\tau_{BH} \propto e^{-(\Pi_i N_i) \log (\Pi_i N_i)}$. In a very simple theory based on $SU(5)\times SU(7)\times U(1)$ with quarks in the bi-fundamental representation, PBHs could be as light as a few milligrams without evaporating by today.

We have also shown that despite the large-$N$ theories we studied naturally have exponentially long-lived extremal BH solutions, they are in perfect compatibility with the swampland conjectures. They are compatible with completeness of the spectrum as well as the so-called species bound, which imposes a non-trivial upper limit to the number of colours $N\lesssim \MPl/\Lambda_N$. This upper bound to the number of colours is the strongest that we obtained from swampland-like arguments (see also~\cite{Kaplan:2019soo,Kaplan:2020tdz} for comparable bounds using other arguments). However, the suppression factor $e^{-N\log N}$ leads to apparent exponential violations of the WGC, as discussed in section~\ref{sec:swamp}. 

\subsubsection*{Future directions}
This work opens multiple, new directions to explore. Perhaps the most exciting one is to study the holographic duals of the large-$N$ BHs that we described. Another exciting direction is to include the quantum corrections recently developed and applied to BH with standard EM charge in~\cite{Brown:2024ajk} to improve the understanding of the near-extremal large-$N$ BH proposed in this paper. As these corrections tend to slow down the neutral Hawking emission process near extremality, we speculate that including them will only make some of our results stronger (e.g. will enhance the lifetime of the near-extremal large-$N$ BH). 

In the context of PBH DM, it will be important to study a concrete formation mechanism. Different possibilities include the mechanisms discussed in~\cite{Araya:2020tds,Diamond:2021scl}, where baryons could have been trapped in PBH that formed due to large-amplitude, short-wavelength density perturbations in the very early universe, or the mechanism considered in~\cite{Dvali:2021byy} where (neutral) PBH could form from the YM confining dynamics itself. Closely related, it is exciting to consider the high-frequency gravitational wave (GW) signal associated to the formation mechanism of these light compact objects~\cite{Franciolini:2023osw} as well as other implications for early universe cosmology. 

Other interesting directions are to study the possibility of having a non-zero kinetic mixing between the SM hypercharge and the dark $U(1)$ under which the large-$N$ baryons are charged. Indeed, due to such kinetic mixing the baryons could gain a small, effective SM electric charge and could lead to potentially observable signals. Studying other aspects of light BH phenomenology such as lensing, signals from SM matter accretting into the BHs, or high-frequency GW signals associated to merger dynamics (including the effects of the quark corona) are also interesting avenues that we leave for future work.

\section*{Acknowledgments}
We thank Saquib Hassan for collaboration at early stages of this project. 
We thank Prateek Agrawal, Yang Bai, Gabriele Franciolini, Ed Hardy, Junwu Huang, David E. Kaplan, Vazha Loladze, John March-Russell, Miguel Montero and Michele Redi for useful discussions, and Gaurang Ramakant Kane for useful comments on the manuscript.
We also thank Prateek for enlightening conversations about crossing symmetry and the scattering amplitudes of all kinds of extended objects -- from baryons to more complex ones like apples or even the Earth.

\appendix

\section{Small black holes and the validity of the EFT description}
In the light of the results obtained above, it is interesting to perform the following \textit{gedanken} experiment. Let us consider a big BH with the charge of a single baryon. For large enough masses, the BH is far away from extremality, $M_{\rm BH}\gg Q_{\rm BH}M_{P}$. The electric field at the horizon is also extremely small, $eE\ll \Lambda_N^2$. As we have shown thoughout the paper, the BH will preferentially Hawking radiate neutral particles including mesons, photons and gravitons. The emission of charge, that is the single baryon that fell to the BH, is suppressed by a factor $\sim e^{-N\log N}$. This implies that the BH flows horizontally in the $(M_{\rm BH},Q_{\rm BH})$ plane and will approach the extremal condition, $M_{\rm BH}=\sqrt{2}Q_{\rm BH} M_{\rm Pl}$. Strikingly, well before the BH becomes extremal, our  description breaks down. 

As the BH loses mass, it becomes smaller and the electric field grows, $E\sim N/R_{\rm BH}^2$. At some point, the electric field becomes larger than the string tension and the quark corona starts to form, see sec.~\ref{sec:including_gravity}. The pair production of quarks is slow at the beginning, but as the BH evolves the electric field grows and the Schwinger action decreases. The formation of the corona accelerates and at some point the string self-interactions cannot be neglected, as discussed in sec.~\ref{sec:dilute_string_gas}. This occurs when the corona charge is $Q_{\rm cor}\gtrsim \frac{1}{e}\frac{Q_{\rm BH}^2}{Q_{\rm crit}}$. 

In addition to self-interactions between strings, our description of the BH in terms of classical gravity also breaks down eventually. This fact can be seen by estimating the size of the extremal BH with the charge associated to a single baryon. In this case, $R_{\rm BH}^*\sim GM^*_{\rm BH}$ with  $M^*_{\rm BH}=Q^*_{\rm BH} \MPl=\sqrt{N}\MPl$. Using the species bound~\eqref{eq:species_bound},  $\MPl^2\sim N^2\Lambda_{\rm UV}^2$, we obtain the relation 
\begin{equation}
    R^{*}\sim \frac{1}{\sqrt{N}}\frac{1}{\Lambda_{UV}}\,.
\end{equation}
We find that the radius of such extremal BH is parametrically below the length scale associated to the cut-off of the EFT description. 

It is somewhat surprising that the description of a BH with the electric charge of a single baryon seems to require an understanding of string self-interactions and the inclusion of quantum gravity effects even before reaching extremality. A logical possibility that would avoid the breakdown of the EFT, however, is that the process of corona formation together with the string self-interactions becoming relevant could trigger the spontaneous conversion of the BH back to a baryon. If this process happened, one could imagine that before reaching a regime where quantum gravity effects become relevant the BH forms a corona, expels all the quarks, and becomes a baryon again.

\bibliographystyle{utphys}
\bibliography{newrefs_axion-4.bib}

\end{document}